\documentclass[11pt]{article}
\def\lb{\label}
\def\bb{\bibitem}
\def\be{\begin{equation}}
\def\ee{\end{equation}}
\def\ba{\begin{eqnarray}}
\def\ea{\end{eqnarray}}

\def\e{{\rm e}}

\begin{document}
\begin{titlepage}
\title{\begin{flushright}\begin{small}    LAPTH-002/14
\end{small} \end{flushright} \vspace{2cm}
Comment on `Critical scalar field collapse in AdS$_3$: an analytical
approach'}
\author
{G\'erard Cl\'ement\thanks{Email:gerard.clement@lapth.cnrs.fr}\\
\small{LAPTh, Universit\'e de Savoie, CNRS, 9 chemin de Bellevue,} \\
\small{BP 110, F-74941 Annecy-le-Vieux cedex, France} \\ Alessandro
Fabbri\thanks{Email:afabbri@ific.uv.es}\\ \small{Centro Studi e Ricerche
Enrico Fermiâ Piazza del Viminale 1, 00184 Roma, Italy}
\\ \small{Dipartimento di Fisica dell'Universit\`a di Bologna,
Via Irnerio 46, 40126 Bologna, Italy}\\ \small{Dep. de F\'isica
Te\'orica and IFIC, Universidad de Valencia-CSIC,}\\ \small{C. Dr.
Moliner 50, 46100 Burjassot, Spain} }

\maketitle

\abstract{We comment on the derivation of an analytical solution
presented in \cite{BST}, show that it belongs to a family of
separable solutions previously constructed in \cite{crit2}, and
question its relevance to critical collapse.}

\end{titlepage}\setcounter{page}{2}

In a recent paper\cite{BST}, an analytical solution (thereafter
referred to as the BST solution), depending on a real parameter $p$,
to the Einstein field equations for a massless scalar field
collapsing in $AdS_3$ was presented, and argued to be relevant to
critical collapse in the range $p > 1$. We show here that although
the BST solution is indeed a solution of the field equations, its
derivation in \cite{BST} is incorrect for $p > 1$, and that the BST
solution belongs to a larger family of separable solutions to the
field equations presented in \cite{crit2}. We also comment on the
relevance of the BST solution to critical scalar field collapse in
$AdS_3$.

The authors of \cite{BST} parameterize the spacetime metric in
double-null coordinates by
 \be\lb{met}
ds^2 = -e^{2\sigma(u,v)}dudv + r(u,v)^2d\theta^2\,,
 \ee
and make a self-similar ansatz which leads to the master
differential equation for an auxiliary function $y$ depending on the
variable $x=-(\alpha/2\sigma)\ln{\eta}$, where $\eta= u/v$ and
$\alpha$ and $\sigma$ are real constants,
 \be\lb{mast}
\left(\frac{dy}{dx}\right)^2 - \sigma^2y^2 - \frac{\sigma^2}2y^4 =
2E\,,
 \ee
with $E$ a real integration constant. This can be solved in terms of
elliptic functions. Putting $p = \sqrt{1-4E/\sigma^2}$, the authors
of \cite{BST} show that the resulting metric develops an apparent
horizon for $p>1$, i.e. $E < 0$, and argue that the corresponding
critical exponent is $\gamma = 1/2$.

As presented in \cite{BST}, the derivation of the BST solution is
formally incorrect for $p > 1$. From Eq. (29) of \cite{BST}, the
constant $E$ is related to another integration constant $\tilde{c}$
by
 \be
\frac{\tilde{c}}2 =
\frac{\alpha^4}{A^2}\left(\frac{E}{4\sigma^2}\right)\,.
 \ee
The constant $\tilde{c}$ is introduced in the first integral
relating two auxiliary functions $f(\eta)$ and $\rho(\eta)$,
 \be
\e^{2\rho(\eta)} = \frac{\tilde{c}f^2(\eta)}{\eta^{\alpha+1}}
 \ee
(Eq. (23) of \cite{BST}). This shows clearly that $\tilde{c}$, and
thus also $E$, is positive definite, so that only the range $p<1$ is
allowed. Notwithstanding this, the metric (39) of \cite{BST}
 \be\lb{bst}
ds^2 = -\frac{\alpha^2}2(1-p^2)\frac{r^2}{(uv)^{\alpha+1}}\,dudv +
r^2d\theta^2\,, \quad r(u,v) = \frac{(uv)^{\alpha/2}}y\,,
 \ee
with $\phi(u,v)$ given by (12) and (20), is indeed a solution of the
cosmological Einstein-scalar field equations for all real values of
$p^2$, including $p^2>1$ (and $p^2<0$). To check this, it is enough
to replace in (\ref{met}) $-e^{2\sigma}$ by $e^{2\sigma}$, and
follow again the steps of the derivation in \cite{BST}, leading to
the solution (\ref{bst}) with $p^2>1$.

Now we show that the analytical solution of \cite{BST} actually
belongs to the larger family of separable
solutions\footnote{Separable solutions in the context of critical
collapse were also considered in the four-dimensional case in
\cite{husain}.} to the field equations presented in \cite{crit2}. It
was shown in \cite{crit2} that the ansatz
 \be\lb{met2}
ds^2 = F^2(T)\left[- dT^2 + dR^2 + G^2(R)d\theta^2\right]\,, \quad
\phi = \phi(T)
 \ee
leads to a solution of the field equations, provided the functions
$G(R)$, $F(T)$ and $\phi(T)$ solve the differential equations
 \ba
&& G^{'2} - kG^2 = a\,, \lb{first2}\\
&& \dot{F}^{2} - kF^2 = \Lambda F^4 + b^2\,, \lb{mast2}\\
&& \dot{\phi} = \sqrt2bF^{-1}\,, \lb{phi2}
 \ea
where $\dot{} = \partial/\partial_T$, $' =\partial/\partial_R$, and
$k$, $a$ and the scalar field strength $b$ are real integration
constants.

Equation (\ref{mast2}) has the same form as Eq. (\ref{mast}). If we
choose the special solution of Eq. (\ref{first2}),
 \be\lb{ss}
G(R) = \e^{\nu R} \qquad (k = \nu^2\,,\; a = 0)\,,
 \ee
and transform the coordinates $T$ and $R$ and the function $F(T)$ to
double-null coordinates $u$ and $v$ and a function $y(x)$ defined by
 \ba
\e^{\nu R} &=& \frac{\nu}{\sqrt2b}\,(uv)^{\alpha/2}\,, \quad
\e^{\nu T} = \left(\frac{u}{v}\right)^{\alpha/2}\,, \\
F(T) &=& \frac{\sqrt2b}{\nu y(x)} \qquad (x = -(\nu/\sigma)T)\,,
 \ea
Eq. (\ref{mast2}) with $\Lambda = - 1$ goes over into Eq.
(\ref{mast}) with
 \be\lb{Eb}
E=-b^2\sigma^2/\nu^4\,,
 \ee
and the solution (\ref{met2}) goes over into the $p>1$ BST solution
(\ref{bst}). For completeness, we note that the $p<1$ BST solution
may similarly be obtained from the separable ansatz dual to
(\ref{met2})
 \be\lb{met1}
ds^2 = G^2(R)\left[- dT^2 + dR^2 + F^2(T)d\theta^2\right]\,, \quad
\phi = \phi(R)\,,
 \ee
the functions $F$, $G$ and $\phi$ solving the differential equations
 \ba
&& \dot{F}^{2} - kF^2 = a\,, \lb{first1}\\
&& G^{'2} - kG^2 = -\Lambda G^4 + b^2\,, \lb{mast1}\\
&& \phi' = \sqrt2bG^{-1}\,, \lb{phi1}\,.
 \ea

Finally, we comment on the relevance of the BST solution to critical
collapse. The authors of \cite{BST} show that the solution
(\ref{bst}) with $p^2>1$ has an apparent horizon with a mass aspect
 \be\lb{scale}
M_{AH} = r_{AH}^2 \propto \sqrt{p^2-1}\,,
 \ee
while there is no apparent horizon for $p^2<1$. They conclude that
the critical value of the parameter $p$ is $p=1$, corresponding to
$E=0$, and the mass critical exponent is $1/2$. As $E=0$ implies,
from Eq. (\ref{Eb}), that the scalar field strength $b$ vanishes,
this would mean that the critical solution for scalar field collapse
is a vacuum solution, and thus is constant curvature. This
contradicts the findings of the numerical simulations of \cite{PC}
and \cite{HO}, which show a critical regime for a finite critical
value of the scalar field amplitude, with a strong spacelike
curvature singularity.

Concerning the critical solution for scalar field collapse in
$AdS_3$, let us recall that Garfinkle \cite{G} constructed a family
of continuously self-similar solutions of the $\Lambda=0$ equations,
depending on an integer $n$, and found that the $n=4$ solution was a
good fit to the numerical data of \cite{PC} for critical collapse
near the singularity (where the effect of the cosmological constant
is negligible). Garfinkle and Gundlach \cite{GG} performed the
linear perturbation analysis of the Garfinkle solutions, and showed
that with suitable boundary conditions the $n=2$ Garfinkle solution
admitted a single growing mode, suggesting that this solution should
be the critical one near the singularity. In \cite{CCF}, the
Garfinkle solutions were extended to solutions of the full field
equations truncated to first order in the cosmological constant
$\Lambda$, and the zeroth order linear perturbation analysis of
\cite{GG} was extended to the same order, confirming the results of
that analysis. Thus it would seem that, insofar as the boundary
conditions used in \cite{GG} and \cite{CCF} are appropriate, the
critical solution for scalar field collapse in $AdS_3$ is an
extension of the $n=2$ Garfinkle solution.

\end{document}